\documentstyle[sprocl,epsfig]{article}

\def\st{\scriptstyle}

\def\be{\begin{equation}}
\def\ee{\end{equation}}
\def\bea{\begin{eqnarray}}
\def\eea{\end{eqnarray}}

\newskip\humongous \humongous=0pt plus 1000pt minus 1000pt
\def\caja{\mathsurround=0pt}
\def\eqalign#1{\,\vcenter{\openup1\jot \caja
        \ialign{\strut \hfil$\displaystyle{##}$&$
        \displaystyle{{}##}$\hfil\crcr#1\crcr}}\,}
\newif\ifdtup

\def\eqright #1\cr{\noalign{\hfill$\displaystyle{{}#1}$}}
\def\eqleft #1\cr{\noalign{\noindent$\displaystyle{{}#1}$\hfill}}

\def\oldreffmt#1{\rlap{[#1]} \hbox to 2\parindent{}}

\def\figfmt#1{\rlap{Figure {#1}} \hbox to 1in{}}

\def\begineq #1\endeq{$$ \refstepcounter{equation}\eqalign{#1}\eqno
	(\theequation) $$}
\def\contlimit{\,{\hbox{$\longrightarrow$}\kern-1.8em\lower1ex
\hbox{${\scriptstyle (a\rightarrow0)}$}}\,}
\def\centeron#1#2{{\setbox0=\hbox{#1}\setbox1=\hbox{#2}\ifdim
\wd1>\wd0\kern.5\wd1\kern-.5\wd0\fi
\copy0\kern-.5\wd0\kern-.5\wd1\copy1\ifdim\wd0>\wd1
\kern.5\wd0\kern-.5\wd1\fi}}
\def\centerover#1#2{\centeron{#1}{\setbox0=\hbox{#1}\setbox
1=\hbox{#2}\raise\ht0\hbox{\raise\dp1\hbox{\copy1}}}}
\def\centerunder#1#2{\centeron{#1}{\setbox0=\hbox{#1}\setbox
1=\hbox{#2}\lower\dp0\hbox{\lower\ht1\hbox{\copy1}}}}
\def\lsim{\;\centeron{\raise.35ex\hbox{$<$}}{\lower.65ex\hbox
{$\sim$}}\;}
\def\gsim{\;\centeron{\raise.35ex\hbox{$>$}}{\lower.65ex\hbox
{$\sim$}}\;}
\def\st#1{\centeron{$#1$}{$/$}}

\def\super#1{\ifmmode \hbox{\textsuper{#1}}\else\textsuper{#1}\fi}
\def\textsuper#1{\newcount\holdspacefactor\holdspacefactor=\spacefactor
$^{#1}$\spacefactor=\holdspacefactor}

\def\getcite#1,{\advance\citenumber by1
\ifnum\citenumber=1
\ref{#1}\let\next=\getcite\else\ifx#1@\let\next=\relax
\else ,\ref{#1}\let\next=\getcite\fi\fi\next}

\def\upon #1/#2 {{\textstyle{#1\over #2}}}
\relax

\def\til#1{\centeron{\hbox{$#1$}}{\lower 2ex\hbox{$\char'176$}}}
\def\tild#1{\centeron{\hbox{$\,#1$}}{\lower 2.5ex\hbox{$\char'176$}}}
\def\sumtil{\centeron{\hbox{$\displaystyle\sum$}}{\lower
-1.5ex\hbox{$\widetilde{\phantom{xx}}$}}}

\def\kbar{\underline{k}}

\def\pom{{\rm P\kern -0.53em\llap I\,}}
\def\spom{{\rm P\kern -0.36em\llap \small I\,}}
\def\sspom{{\rm P\kern -0.33em\llap \footnotesize I\,}}

\begin{document} 

\begin{titlepage} 

\rightline{\vbox{\halign{&#\hfil\cr
&ANL-HEP-CP-98-130
 \cr
&\today\cr}}} 
\vspace{.75in} 

\begin{center}
{\bf SOLVING QCD VIA MULTI-REGGE THEORY}\footnote{Work 
supported by the U.S.
Department of Energy, Division of High Energy Physics, \newline Contracts
W-31-109-ENG-38 and DEFG05-86-ER-40272} 
\medskip

Alan. R. White\footnote{arw@hep.anl.gov }
\end{center}
\vskip 0.6cm

\centerline{High Energy Physics Division}
\centerline{Argonne National Laboratory}
\centerline{9700 South Cass, Il 60439, USA.}
\vspace{0.5cm}

\begin{abstract} 

A high-energy, transverse momentum cut-off, solution
of QCD is outlined. Regge pole and ``single gluon'' properties of 
the pomeron are directly related to the confinement and chiral symmetry 
breaking properties of the hadron spectrum. 
This solution, which corresponds to a supercritical phase
of Reggeon Field Theory, 
may only be applicable to QCD with a very special quark 
content. 

\end{abstract} 

\vspace{1.5in}
\begin{center}

Invited talk presented at the ``4th Workshop on Quantum Chromodynamics'',
\newline The American University of Paris, Paris France, June 1-6, 1998.

\end{center}

\end{titlepage}

\section{Introduction}

The QCD pomeron is usually discussed without much attention paid to
the scattering states. States containing only elementary constituents are 
normally considered. As a matter of principle, a full solution
of QCD at high-energy requires that we find both the true hadronic 
states and the exchanged pomeron giving scattering amplitudes.
Unitarity must be satisfied in both the $s$-channel and the 
$t$-channel. 

Experimentally the pomeron appears, approximately, to be a Regge
pole at small $Q^2$ and~\cite{h1} a single gluon at larger $Q^2$. Neither
property is present in QCD perturbation theory. In the 
high-energy, transverse momentum cut-off, ``solution'' of QCD that I 
outline in this talk the experimental ``non-perturbative properties'' of the
pomeron are directly related to the confinement and chiral symmetry breaking
properties of hadrons. That is, experimental properties of the pomeron 
are directly related to special properties of the scattering states. 
It is particularly interesting that 
our solution may only be applicable to QCD with a very special quark 
content.\footnote{In general our work leads us to 
doubt very strongly that ``We now know that there are an infinite number of 
consistent S-Matrices that satisfy all the sacred principles. ... any 
gauge group, and many sets of fermions''~\cite{dg}} 

Our arguments involve the techniques of multi-regge QCD calculations, 
the dynamics of the massless quark U(1) anomaly, which
will be the main focus of this talk, and reggeon field theory
phase-transition analysis, which we will avoid almost completely. Our
results show~\cite{arw97} how confinement and chiral
symmetry breaking, normally understood as consequences of the vacuum, can 
instead be produced by a ``wee-parton'' distribution. This is a very
non-trivial property that provides, I hope, a deeper basis for the parton
model (and even~\cite{kw} the constituent quark model) in QCD ! 

Multi-Regge Theory is an abstract formalism~\cite{arw97}
that we developed a major part 
of in the 70's. The formalism is based on the existence of
asymptotic analyticity domains for multiparticle amplitudes
derived~\cite{arw1,sw} 
via ``Axiomatic
Field Theory'' and ``Axiomatic S-Matrix Theory''. All the assumptions made
are expected to be valid in a completely massive spontaneously-broken gauge
theory. Since we begin with massive reggeizing gluons, this is
effectively the starting point for our analysis of QCD. 

For our purposes, the most important component of multi-regge theory 
is the reggeon unitarity equations~\cite{arw1,gpt}. Using these equations, 
well-known Regge limit QCD
calculations~\cite{bfkl,bs,fs1} can be extended to obtain 
amplitudes, of the form illustrated in Fig.~1, involving multiple exchanges of 
reggeized gluons
and quarks in a variety of channels. The central idea of our work is that we 
can, by
considering  
infra-red limits in which
gluon mass(es) $M$ and quark mass(es) 
$m ~ \to 0$, find hadrons (e.g. the pion) and the $~\pom~$ 
as (coupled) Regge pole bound states of 
reggeons. Presently the simultaneous study of 
bound states and their scattering amplitudes is impossible in any other
formalism. 

\noindent \parbox{3in}{
\leavevmode
\epsfxsize=2.7in
\epsffile{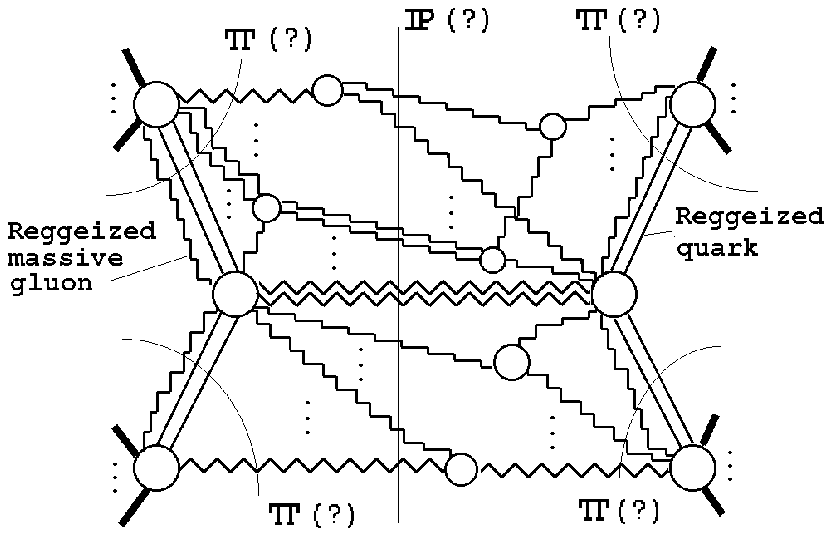}
}
\parbox{1.7in}{
\leavevmode
\epsfxsize=1.5in
\epsffile{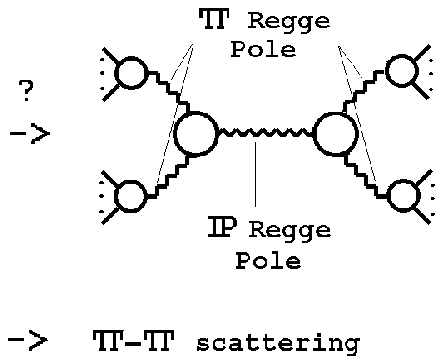}
}

\vspace{0.1in}

\begin{center}
Fig.~1 The Anticipated Formation of Pion 
Scattering Amplitudes
\end{center}

In general, limits wrt all mass, gauge symmetry, and cut-off parameters
are crucial. The most important feature, however, is the dynamical role 
played by new ``reggeon helicity-flip'' vertices that appear in the amplitudes
we discuss. The hadron amplitudes we obtain are initially isolated via a
(``volume'') infra-red divergence that appears when SU(3) gauge symmetry is
partially broken to SU(2) and the 
limit of zero quark mass is also taken. The divergence involves
quark loop helicity-flip vertices containing chirality
violation (c.f. instanton interactions). 
The chirality violation survives the massless quark limit 
because of an infra-red effect closely related to 
the triangle anomaly~\cite{cg}. The divergence produces a ``wee parton
condensate'' that is directly responsible, while the gauge
symmetry is partially broken, for confinement and chiral symmetry breaking. 
The pomeron is a reggeized gluon in 
the wee
parton condensate and so is obviously a Regge pole. We will not give 
a description of the supercritical pomeron~\cite{arw1} in this talk, however, 
all the essential features of this RFT phase are present in the solution of 
partially-broken QCD that we present.

We discuss the restoration of SU(3) gauge symmetry only briefly. The 
increase in the gauge symmetry is closely related to 
the critical behaviour of the pomeron~\cite{cri} 
and the associated disappearance of the supercritical condensate. We note 
that the large $Q^2$ of deep-inelastic scattering provides a 
finite volume constraint that can keep the theory (locally) in the
supercritical RFT phase as the full gauge symmetry is restored. A single gluon
(in the background wee parton condensate) should then be a good
approximation for the pomeron. Finally we discuss the 
circumstances under which our solution can be realized in QCD. 

\section{Reggeon Diagrams in QCD}

Leading-log Regge limit calculations of elastic and multi-regge production 
amplitudes in (spontaneously-broken) gauge theories 
show\cite{bfkl,bs,fs1} that both gluons and quarks ``reggeize'', i.e.
they lie on Regge trajectories. Non-leading log calculations are described
by ``reggeon diagrams'' involving reggeized gluons and
quarks. Reggeon unitarity implies that a complete set of reggeon
diagrams arise from higher-order contributions.

Gluon reggeon diagrams involve a reggeon propagator for each reggeon state 
and also gluon particle poles e.g. the two-reggeon state 
\newline\parbox{0.8in}{
\begin{center}
\leavevmode
\epsfxsize=0.5in
\epsffile{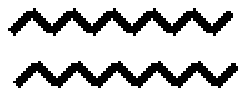}
\end{center}
}
\parbox{3.9in}{
$$
~\longleftrightarrow ~~~
\int {d^2k_1 \over (k_1^2 +M^2) } {d^2k_2 \over (k_2^2 + M^2)}~ 
{\delta^2(k_1'+k_2'-k_1-k_2)
\over J-1 +  \Delta(k_1^2) + \Delta(k_2^2)}
$$}
\newline The BFKL equation~\cite{bfkl}
corresponds to 2-reggeon unitarity i.e. iteration of the 2-reggeon state 
with a 2-2 reggeon interaction $ R_{22}= 
[(\kbar^2_1+M^2)({\kbar^2_2}'+M^2)+(\kbar^2_2+M^2)(
{\kbar^2_1}'+M^2)]/[ (\kbar_1-\kbar_1')^2+M^2] ~+ \cdots 
$

We will be interested in the limit $M \to 0$ and will assume that two 
leading-order properties of reggeon diagrams 
generalize to all orders. The first is that infra-red divergences 
exponentiate to zero all diagrams that do not carry
zero color in the $t$-channel. The second property is that 
infra-red finiteness implies canonical 
scaling ($\sim Q^{-2}$) for color zero reggeon 
amplitudes when all transverse momenta are simultaneously scaled to zero
(this requires $\alpha_s(Q^2) \st{\to} \infty$ when $Q^2 \to 0$).

\section{Reggeon Diagrams for Helicity-Pole Limit Amplitudes}

The generality of reggeon unitarity is most powerful when 
applied to the amplitudes appearing in ``maximal helicity-pole limits''. 
Amplitudes of this kind contain the bound-state reggeon scattering 
amplitudes we are looking for and, since the appropriate 
Sommerfeld-Watson representation shows that only a single
(analytically-continued) partial-wave amplitude is involved, 
reggeon unitarity implies that such limits can be described by reggeon 
diagrams. (The physical 
significance of such diagrams is
subtle~\cite{arw97} in that ``physical'' $k_{\perp}$ planes 
in general contain lightlike momenta !) 

As an example, we consider a maximal helicity-flip limit 
for an 8-pt amplitude. We introduce angular variables variables~\cite{arw97} 
as illustrated in Fig.~2.
We consider the ``helicity-flip'' limit 
$ z,u_1,u^{-1}_2,u_3,u^{-1}_4 \to \infty $. 
The behavior of invariants in this limit is
\newline \parbox{2.3in}{\openup\jot
$$
\eqalign{&P_1.P_2 \sim u_1u^{-1}_2~, ~~~P_1.P_3 \sim u_1zu_3~, \cr
& P_2.P_4 \sim u^{-1}_2u^{-1}_4~, ~~~P_1.Q_3 \sim u_1z~, \cr
&Q_1.Q_3 \sim z~, ~~~P_4.Q_1 \sim zu^{-1}_4 
~ ~~ \cdots \cr
&~ P_1.Q,~P_2.Q,~ P_3.Q,~P_4.Q  ~~~\hbox{{\it finite }} }
$$
($u_1,u_2^{-1} \to \infty$ is a ``helicity-flip'' limit,
$u_1,u_2 \to \infty$ is a ``non-flip'' limit.)
}
\parbox{2.4in}{
\begin{center}
\leavevmode
\epsfxsize=2.1in
\epsffile{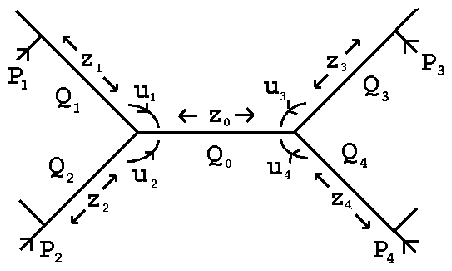}
\newline Fig.~2 Variables for the 8-pt 
\newline Amplitude
\end{center}}

\vspace{0.1in}

Reggeon unitarity determines that the helicity-flip
limit is described by 
\newline \parbox{2.4in}{ 
reggeon diagrams of the form shown in Fig.~3. The amplitudes 
$~{\raisebox{-2mm}{\epsfxsize=0.3in \epsffile{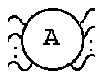}}}~$
contain all elastic scattering 
reggeon diagrams. The $T^F$ are new ``reggeon helicity-flip'' vertices
that play a crucial role in our QCD analysis. (These vertices do not appear 
in elastic scatttering reggeon diagrams).}
\parbox{2.3in}{
\begin{center}
\leavevmode
\epsfxsize=2in
\epsffile{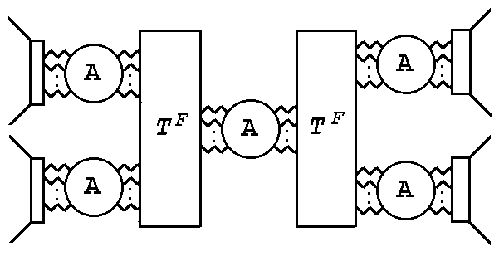}

Fig.~3 Reggeon Diagrams for 
\newline the 8-pt Amplitude
\end{center}
}

\section{Reggeon Helicity-Flip Vertices }

The $T^F$ vertices are most simply isolated kinematically by considering
a ``non-planar'' triple-regge limit which, for simplicity, we will define by 
introducing three distinct light-cone momenta. (This limit
actually gives a sum of
three $T^F$ vertices of the kind discussed above~\cite{arw97}, but in this
talk we will not elaborate on this subtlety.)  We 
use the tree diagram of Fig.~4(a) to define momenta and study the special 
kinematics 
\newline \parbox{2in}{
$$
\eqalign{&P_1\to (p_1,p_1,0,0),~~~p_1 \to \infty \cr
&P_2\to (p_2,0,p_2,0),~~~p_2 \to \infty \cr
&P_3\to(p_3,0,0,p_3),~~~p_3 \to \infty  \cr
&~ \cr
&Q_1\to (0,0,q_2, -q_3) \cr
&Q_2\to (0,-q_1,0,q_3) \cr
&Q_3\to (0,q_1,-q_2,0) }
$$
}
\parbox{2.7in}{
\begin{center}
\leavevmode
\epsfxsize=2.5in
\epsffile{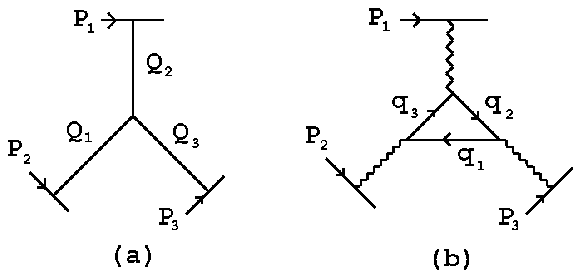}
\newline Fig.~4 (a) A Tree Diagram and (b) a quark loop coupling for three 
quark scattering.
\end{center}
}

Consider, first, three quarks scattering
via gluon exchange with a quark loop coupling as in Fig.~4(b). The 
non-planar triple-regge limit gives  
$$
\to
~ g^6~~ { p_1p_2p_3 \over t_1 t_2 t_3 } ~~\Gamma_{1^+2^+3^+}(q_1,q_2,q_3)
~~~\leftrightarrow ~~~ 
g^3~~ { p_1p_2p_3 \over t_1 t_2 t_3 }~ T^F(Q_1,Q_2,Q_3) 
$$ 
where $~ \gamma_{i^+} = \gamma_0 + \gamma_i $ and 
$\Gamma_{\mu_1 \mu_2 \mu_3}$ is given by the quark triangle diagram i.e. 
$$
\Gamma_{\mu_1 \mu_2 \mu_3} = i\int {  d^4 k~ Tr \{ \gamma_{\mu_1}
(\st{q}_3 + \st{k} + m ) \gamma_{\mu_2} (\st{q_1} + \st{k} + m ) 
\gamma_{\mu_3} (\st{q}_2 + \st{k} + m) \} 
\over [ (q_1 + k)^2 - m^2 ][ (q_2 + k)^2 - m^2 ]
[ (q_3 + k)^2 - m^2 ]}
$$
where $m$ is the quark mass. We denote the $O(m^2)$ chirality-violating part of 
$~T^F ~(\equiv ~g^3 ~\Gamma_{1^+2^+3^+}~)$ by 
$T^{F,m^2}~$ and note that the 
limits $q_1, q_2, q_3 \sim Q \to 0$ and $m \to 0$ do not commute, i.e. 
$$
T^{F,m^2} {\centerunder{$\sim$}{\raisebox{-5mm} 
{$Q \to 0$} }}~T^F_0~ =~ Q ~i~m^2 \int {d^4k \over [ k^2 - m^2 ]^3 }
 ~~~~= ~  R ~Q 
$$
where $R$ is independent of $m$. This non-commutativity is an ``infra-red 
anomaly'' due to the triangle Landau singularity~\cite{cg}.

\section{The Infra-Red Anomaly in Helicity-Flip Vertices}

After color factors are included and all related diagrams summed, $T^F_0$
survives only in very special vertices coupling reggeon states
with ``anomalous color parity''. We define color parity ($C_c$) via the
transformation $ A^i_{ab} \to - A^i_{ba}$ 
for gluon color matrices and say that a reggeon state has anomalous color 
parity if the signature $\tau$ (i.e. whether the number of 
reggeons is even or odd) is not equal to the color parity. 

In general, the $T^F$ vertices coupling multiple-reggeon states contain many 
reduced quark loops. The vertices containing $T^F_0$ also contain 
ultra-violet divergences associated with the anomaly. To maintain the
reggeon Ward identities that ensure gauge invariance~\cite{arw97}, we
introduce Pauli-Villars fermions as a regularization. (Note that we take the
regulator mass $m_{\Lambda} \to \infty$ after $m \to 0$. This implies that
the initial 
theory with $m \neq 0$ is non-unitary for $k_{\perp} \gsim m_{\Lambda}$.) 
The regulated vertex, $T^{{\cal F},m^2}$, satisfies 
$$
T^{{\cal F},m^2}(Q) \sim ~
T^{F,m^2} - T^{F,m_{\Lambda}^2}
~{\centerunder{$\sim$}{\raisebox{-4mm} 
{$Q \to 0$} }} Q^2 ~, ~~  T^{{\cal F},0}(Q) 
{\centerunder{$\sim$}{\raisebox{-4mm} 
{$Q \to 0$} }}
T^F_0 \sim  ~Q 
$$
implying that 
imposing gauge invariance for $m \neq 0$ gives a 
slower vanishing as $Q \to 0$ when $m=0$.

We will be particularly interested
in the ``anomalous odderon''
three-reggeon 
\newline \parbox{2.2in}{ state with color factor
$f_{ijk}A^iA^jA^k$ that has $\tau = -1$ but $C_c = +1$ 
\newline (c.f. the winding-number 
current
\newline $K_{\mu}=\epsilon_{\mu \nu \gamma 
\delta}f_{ijk}A^i_{\nu}A^j_{\gamma}A^k_{\delta}$ ). 
\newline $T^{{\cal F},0}(Q)$ 
appears in the triple coupling of three anomalous 
odderon states as in Fig.~5. 
}
\parbox{2.5in}{
\begin{center}
\leavevmode
\epsfxsize=2.2in
\epsffile{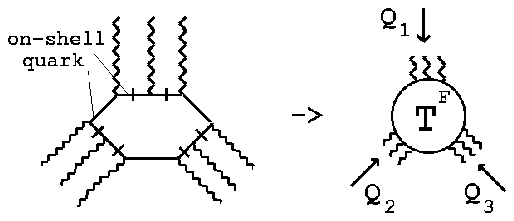}
\newline Fig.~5 An Anomalous Odderon 
\newline Triple Coupling.
\end{center}
}

\section{A Quark Mass Infra-Red Divergence} 

A vital consequence of the ``anomalous'' behavior of $~T^{{\cal F},0}~$ 
as $~Q \to 0~$ is that an 
additional infra-red divergence 
is produced (as $m \to 0$) in massless gluon reggeon diagrams. 
The divergence occurs in diagrams involving the $T^F$ 
where $~Q_1 \sim Q_2 \sim Q_3 \sim 0
~$ is part of the integration region. This requires that $T^F$ 
\newline \parbox{2.1in}{ 
be a 
disconnected component of a vertex coupling 
distinct reggeon channels, 
as in Fig.~6. 
In this diagram an anomalous odderon reggeon state ($~\equiv~$ 
$~\raisebox{-1.5mm}{$\epsfxsize=0.4in 
\epsffile{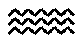}$}~$) is denoted by
$~\raisebox{-0.5mm}{$\epsfxsize=0.3in \epsffile{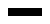}$}~$
while 
$~\raisebox{-0.5mm}{$\epsfxsize=0.3in \epsffile{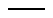}$}~$
denotes any normal reggeon state. Fig.~6 is of the general form illustrated 
in Fig.~1.}
\parbox{2.6in}{ 
\begin{center}
\leavevmode
\epsfxsize=2.3in 
\epsffile{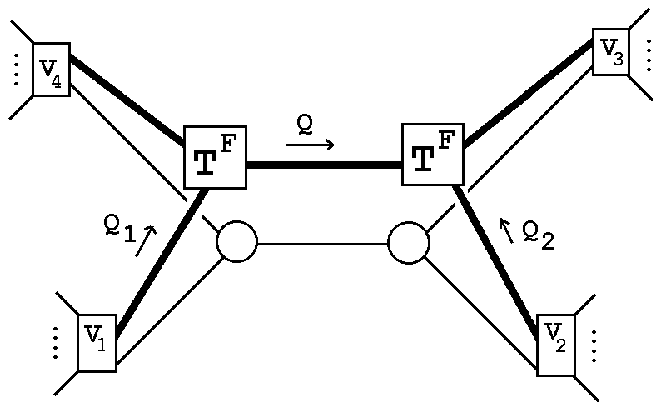}
\newline Fig.~6 A Divergent Reggeon Diagram 
\end{center}
}

The canonical 
scaling of the anomalous odderon states 
gives the infra-red behaviour 
$$
\eqalign{& \int \cdots {d^2Q_1 ~ d^2Q_2 
~d^2Q \over Q_1^2 Q_2^2 Q^2 (Q-Q_1)^2 (Q - Q_2)^2 }
~~V_1(Q_1)V_2(Q_2) V_3(Q- Q_2) V_4(Q - Q_1) \cr
&\times  ~ T^{ {\cal F}}(Q_1,Q)
T^{ {\cal F}}(Q,Q_2)
~\times~\hbox{[regular vertices and reggeon propagators]}
}
$$
for Fig.~6. Depending on the behaviour of the $V_i~$, it is clear that a
divergence may indeed occur when $Q \sim Q_1 \sim Q_2 \to 0$.

The divergence of Fig.~6 is preserved and a possible 
cancelation~\cite{arw97} eliminated 
if we partially break the SU(3) gauge symmetry to SU(2). 
In this case, a divergence can occur in any diagram of the form 
of Fig.~8 in which 
$~\raisebox{-0.5mm}{$\epsfxsize=0.3in \epsffile{dss15.ps}$}~$
is any SU(2) singlet 
combination of massless gluons with 
$~C_c= -\tau = +1~ $ (i.e. a generalized SU(2) anomalous odderon) and  
$~\raisebox{-0.5mm}{$\epsfxsize=0.3in \epsffile{dss16.ps}$}~$
is any normal reggeon state 
containing one or more SU(2) singlet 
massive reggeized gluons (or quarks). 

A-priori reggeon Ward identities imply $ V_i \sim Q_i$ when 
$Q_i \to 0, ~\forall ~ i $, which would be sufficient to eliminate any
divergence in Fig.~6. However, if we impose the ``initial condition'' that 
$V_1,V_2 ~ \st{\rightarrow}~ 0$, the divergence is present 
in a general class of diagrams, including 
those having the general structure illustrated in 
\newline \parbox{2in}{
Fig.~7. In this diagram there are 
$n + 3$ 
multi-reggeon states of the form
$\raisebox{-2mm}{$\epsfxsize=0.4in \epsffile{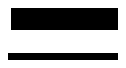}$}$ .
Imposing $V_1,V_2 ~ \st{\rightarrow}~ 0$ and assuming 
that reggeon Ward identities are satisfied by the remaining vertices, i.e. 
$$
V_i(Q_i) \sim V(Q_i) = Q_i 
$$
$i \neq 1,2$, gives that 
Fig.~7 has
}
\parbox{2.7in}{
\begin{center}
\leavevmode
\epsfxsize=2.4in
\epsffile{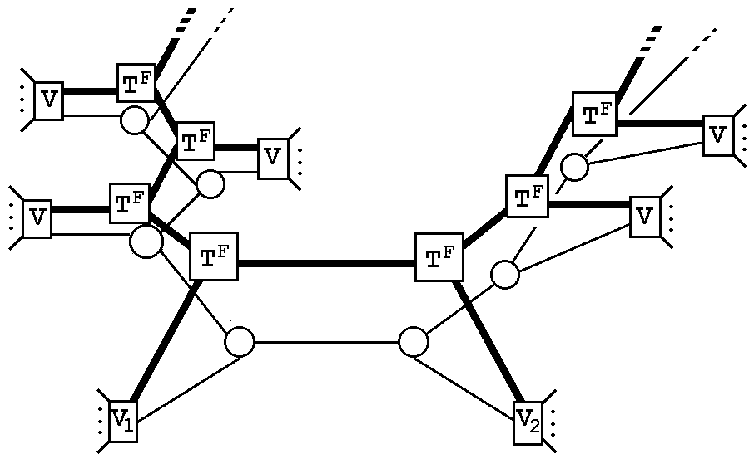}

Fig.~7 A General Divergent Diagram
\end{center}
}
\newline 
the infra-red behavior
$$ 
\int {d^2 Q \over Q^2}~\left[\int {d^2 Q \over Q^4}\right]^n 
~~\left[V(Q)~ T^ {\cal F}(Q)\right]^n 
$$
giving (as $m \to 0$) an overall logarithmic divergence. 
In general, this divergence occurs in just those multi-reggeon 
diagrams which contain only SU(2) color zero states of the form 
$~\raisebox{-2mm}{$\epsfxsize=0.4in \epsffile{cspp12.ps}$}~$
coupled by regular and $~T^{{\cal F},0}~$ vertices, as in the examples we 
have discussed.

\section{Confinement and a Parton Picture}

We define physical amplitudes by extracting the 
coefficient of the logarithmic divergence. There is ``confinement''
in that a particular set of color-zero reggeon states is selected that
contains no massless multigluon states and has the necessary completeness
property to consistently define an S-Matrix. That is, if two or more 
such states scatter 
via QCD interactions, the final states contain only 
arbitrary numbers of the
same set of states. Since $k_{\perp} =0$ for the anomalous odderon component of
each reggeon state, an ``anomalous odderon condensate'' 
is generated. 
The form of physical amplitudes is illustrated in 
\newline \parbox{2in} { Fig.~8. 
In addition to 
the $k_{\perp} =0$ (``wee-parton'') component, 
each physical reggeon state has a 
finite momentum ``normal'' parton component carrying 
the kinematic properties of interactions. 
We emphasize that the ``scattering'' of the $k_{\perp} = 0$ condensate is
directly due to the infra-red quark triangle anomaly. 
}
\parbox{2.7in}{
\begin{center}
\leavevmode
\epsfxsize=2.2in
\epsffile{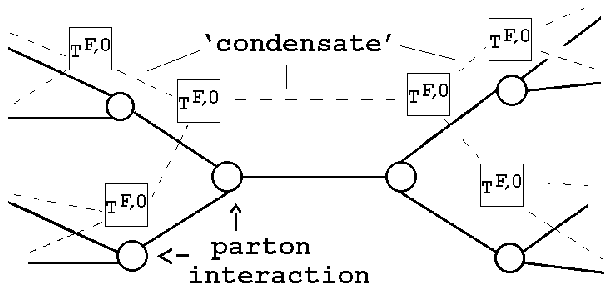}
\newline $~$
\newline Fig.~8 A Physical Amplitude
\end{center}
}

The breaking of the gauge symmetry has produced physical states in which 
the ``partons'' are separated into a universal wee-parton component and a 
normal reggeon parton component which is distinct in each distinct 
physical state. However, the condensate has the important 
property that it switches the signature compared to that of the normal parton 
component. The following are a direct consequence. 

i) The ``pomeron'' has a reggeized gluon normal parton component, but is a 
Regge pole with $\tau  = - C_c = + 1$ and intercept $\neq 0$. 

ii) There is a bound-state reggeon formed from two massive SU(2) doublet 
gluons, giving an exchange-degenerate partner
to the pomeron. The SU(2) singlet massive gluon lies on this trajectory. 

iii) There is chiral symmetry breaking. 
Studies of $\tau = -1$ quark-antiquark exchange~\cite{ks}
to demonstrate a ``reggeization 
cancelation'' generating a Regge pole  
with zero intercept and 
with  
$\tau = -1, ~C_c = +1$ and $P = -1$. In the condensate this gives a Regge 
pole with  
$\tau = +1, ~ P = -1$, resulting in the massless pion associated with chiral 
symmetry breaking. 

Although we have not discussed Reggeon Field Theory, 
we note that all the 
features of my supercritical RFT solution~\cite{arw1} are present.
(This solution was very controversial 20 years ago - although it was
supported by Gribov !) 

\section{Restoration of SU(3) Gauge Symmetry }

We make only a few brief comments on this, obviously important, subject.
Because of complimentarity~\cite{fs},
restoring SU(3) symmetry (decoupling a color triplet Higgs
scalar field) should be straightforward if we impose $k_{\perp} <
\Lambda_{\perp}$. Restoring the symmetry removes the mass scale that 
produces the reggeon condensate. If the (partially) broken theory can be mapped 
completely onto supercritical RFT then the condensate 
and the odd-signature partner for the pomeron will disappear simultaneously
and the critical pomeron~\cite{cri} will be the result. The wee-parton
condensate will be replaced by a universal, small $k_{\perp}$, wee parton,
critical phenomenon that merges smoothly with the large 
$k_{\perp}$ normal (or constituent) parton component of physical states,
just as originally envisaged by Feynman~\cite{rf}. (Note that, because of
the odd SU(3) color charge parity of the pomeron, the two-gluon BFKL pomeron 
will not contribute.) 

RFT implies that 
$\Lambda_{\perp}$ mixes with the symmetry breaking mass scale 
and becomes a ``relevant parameter'' for the
critical behavior. After the symmetry breaking scale is 
removed, there will (for a general number of quark flavors) be a 
$~\Lambda_{\perp c}$ such that 
$\Lambda_{\perp} > \Lambda_{\perp c} ~$ implies the pomeron is in the 
subcritical phase, while $\Lambda_{\perp} <  \Lambda_{\perp c}~ $ will
give the 
supercritical phase.  
This implies that the supercritical phase can be
realized with the full gauge symmetry restored, if $\Lambda_{\perp} $ is taken 
small enough. $\alpha_{\spom}(0)$ will be a function of
$\Lambda_{\perp}$. In deep-inelastic diffraction 
large $Q^2 $ will act as an additional (local) lower $k_{\perp}$ cut-off and
produce a ``finite volume'' effect that can keep the theory supercritical as
the SU(3) symmetry is restored. This implies that the pomeron should 
be well approximated by a single (reggeized) gluon 
(in a soft gluon background) in DIS diffraction. 

To remove $\Lambda_{\perp}$ requires $\Lambda_{\perp c} ~= \infty $. 
This requires a specific quark flavor
content. For any quark content, we can take 
$\Lambda_{\perp} <<  \Lambda_{\perp c} ~  $, and go deep into the 
supercritical phase. We obtain a picture in which constituent quark 
hadrons interact via a massive composite ``gluon'' (and an exchange 
degenerate pomeron). Confinement and chiral symmetry breaking are realized 
via a simple, universal, wee parton component of physical states. This is 
remarkably close to the realization of the constituent quark model
via light-cone quantization that has been advocated by light-cone 
enthusiasts~\cite{kw}.

\section{When is this Solution Realized in QCD ??}

We have found a high-energy S-Matrix via a transverse momentum infra-red
phenomenon involving massless gluons and quarks. At first sight, 
this should not occur in QCD since non-perturbative effects
should eliminate massless gluons for $k_{\perp} < \lambda_{QCD}$! Our
solution requires that massless QCD 
remain weak-coupling at $k_{\perp} = 0$. This is generally anticipated to be 
the case only if there are
a sufficient number of massless quarks in the theory to 
give an infra-red fixed point for $\alpha_s$.

There is such an infra-red fixed point when the number of flavors is the the
maximum allowed by asymptotic freedom. In this case SU(3) symmetry
can be broken to SU(2) with an asymptotically-free scalar field. This 
implies 
that $\Lambda_{\perp c} = \infty $, so that critical 
pomeron scaling occurs for all $k_{\perp}$, allowing a smooth match 
with perturbative QCD. 

The above arguments suggest that if ``single gluon'' supercritical pomeron 
behavior is actually observed at HERA then new QCD physics, in the form 
of a new fermion sector, remains to be discovered above the (diffractive) 
$Q^2$ range presently covered. Everything is consistent if the electroweak
scale is a QCD scale, i.e. the ``Higgs sector'' of the Standard Model, that 
is yet to be discovered, is
composed~\cite{arw2} of higher-color (sextet) quarks. 
A special definition of QCD is necessarily involved, but we will not 
discuss this here.

\noindent { \bf References}

\end{document}